# EVOLUTION OF SCHRÖDINGER UNCERTAINTY RELATION IN QUANTUM MECHANICS

*A. Angelow*

*72 Trackia Blvd., G. Nadjakoff Institute of Solid State Physics, Bulgarian Academy of Sciences*
*a_angelow@phys.bas.bg*

**Abstract.** In the present article, we discuss one of the basic relations of Quantum Mechanics – the Uncertainty Relation (UR). In 1930, few years after Heisenberg, Erwin Schrödinger generalized the famous Uncertainty Relation in Quantum Mechanics, making it more precise than the original. The present study discusses recent generalizations of Schrödinger's work and explains why his paper remains almost forgotten in the last century.

**Key words:** quantum mechanics, Heisenberg uncertainty relation, Schrödinger uncertainty relation

One of the most revolutionary consequences that quantum mechanics bequeathed as a fundamental principle in physics is the refusal of strong determinism. That is why the uncertainty relation (called uncertainty principle in the beginning of quantum mechanics) plays fundamental role in this science.

The uncertainty relation in quantum mechanics was discussed in the special literature as well as in some philosophical books. However, in the most of the articles about quantum mechanics, the term "uncertainty relation" is associated with the Heisenberg uncertainty principle for canonical quantum observables: position *q* and momentum *p* [1]

$$(1) \qquad \delta x \delta p_x \approx \hbar .$$

Heisenberg, and later the Copenhagen group, interpreted it as the impossibility of simultaneous precise measurement of the canonical quantum observables with a precision bigger than the Planck`s constant *"The more precisely is the position determined, the less precisely is the momentum known, and vice versa"* [1]. Heisenberg's uncertainty principle, however, is a special case and it refers to wave packets with Gauss distribution. The proof of the strong inequality was given by Kennard and Weyl [2].

$$(2) \qquad (\Delta x)(\Delta p_x) \geq \frac{\hbar}{2} .$$

Later Robertson [3] generalized the correlation for arbitrary observables *A* and *B* [1], and Dichburn [4] presented the relation between Heisenberg's fluctuations and the mean quadratic deviation $\frac{\delta A}{\sqrt{2}} = \sqrt{\langle A^2 \rangle - \langle A \rangle^2} \equiv \Delta A$. The symbol $\langle \_ \rangle$ means quantum-mechanical mean value.

Schrödinger [5, 6] and Robertson [7] generalized and précised Heisenberg's principle in 1930

$$(3) \qquad (\Delta A)^2 (\Delta B)^2 \geq \left( \frac{\langle AB + BA \rangle}{2} - \langle A \rangle \langle B \rangle \right)^2 + \left| \frac{\langle AB - BA \rangle}{2} \right|^2 ,$$

where a new term was added.

---

[1] For strong definition of quantum mechanical term **observable** (as a positive operator valued measure) see [12].

According to the probability theory, two random variables have three independent statistical moments of second order– the dispersion of two quantities and their covariance [8]. By definition, the covariance is determined by

(4) $$Cov(A,B) \equiv \frac{\langle AB+BA \rangle}{2} - \langle A \rangle \langle B \rangle = Cov(B,A).$$

When $A = B$, then $Cov(A,A) = \langle A^2 \rangle - \langle A \rangle^2 = Var(A) \equiv (\Delta A)^2$, i.e. the variance is a special case of the covariance. If we use the covariance matrix $\sigma_{(A,B)} = \begin{pmatrix} (\Delta A)^2 & Cov(A,B) \\ Cov(B,A) & (\Delta B)^2 \end{pmatrix}$ [8], the Schrödinger's relation (3) can be expressed in a compact [6] and "very elegant form"[9]

(6) $$\det \sigma \geq \left| \frac{AB-BA}{2} \right|^2, \quad \det \sigma \geq \frac{\hbar^2}{4}.$$

The last relation is for canonical variables $\hat{q}$ и $\hat{p}$, where $\hat{q}\hat{p} - \hat{p}\hat{q} = i\hbar$. The proof of Schrödinger Uncertainty Relation is based on Cauchy-Schwarz inequality [5,6], which is a particular case of the Hölder inequality.

Zero covariance is necessary, but not sufficient condition for the independence of two random variables, in the classical as well as in the quantum statistical physics. In case of zero covariance, the sufficient condition is not satisfied because the statistical correlation between the two random variables still exists (postulated by the very Heisenberg's uncertainty principle). By adding new expression (covariance) Schrödinger defines more accurate the statistical correlation of two physical variables, which is stronger now. Indeed in work [13, 32] one can see, that non-zero covariance due to non-linear effect – light in initial coherent state passing through anisotropic waveguide in $LiNbO_3$ with non-linearity $\chi^{(2)}$, was transformed in light with non-zero covariance $(Cov(A, B) = f(\chi^{(2)}) \neq 0)$. In other words, the non-linearity is responsible for the appearance of Cov(A, B) in quantum-statistical treatment of degenerated parametric amplification. This leads to appearance of additional terms in the Hamiltonian, containing the creation and annihilation operators of second power $(\hat{a}^+)^2$ и $\hat{a}^2$ [13, 32]. We observe similar effect of classical non-linearity in other investigations [19] with fully quantum examination of the phenomenon. Because of the non-zero covariance (and mostly because Schrödinger discussed the new term (covariance) in details [5, 6]) we call this states covariance (Schrödinger Covariance States [13]). The notion covariance is most precise in mathematical point of view[2]. We call them Schrödinger's, because they minimize Schrödinger uncertainty relation. They are subclass of all states minimizing Schrödinger uncertainty relation (Schrödinger Minimum Uncertainty States - SMUS) [15]. In early publications, however (see review [17]), almost no one (with small exceptions [13, 19, 30]) pay attention to the fact that the covariance could be non-zero, which leads to inexact conclusions, based on Heisenberg relation only, that the light (in the mentioned states above) can not minimize the uncertainty relation [14]. To stimulate experimental realization of covariance states and to escape rich (sometimes confusing) terminology we think that it is necessary precisely to specify the notion of that states[3]. The general group of Schrödinger minimum uncertainty states consists of three classes: coherent, squeezed and covariance (the last exclude the presence of coherent and squeezed state). The unsuitable extension of the terms (generalized coherent states, generalized squeezed states, generalized correlated states, two-photon coherent state) does not help so much to experimentalists. Our approach is in direction of making the terms of Schrödinger minimum uncertainty states more precise and this classification of covariance, coherent and squeezed states is done in [15], where the proposal of experimental realization of covariance states is given.

---

[2] With this we stay out about existing term in literature in early period of quantum optics (correlated, two-photon, Stoler SU (1, 1), SU (2) et cetera). One non-precise example in the terminology a correlated states of Manko, including states with zero correlation, but are called correlated. Another example are two-photon states of Yuen, which include one-photon Glauber states |α>, but are called two-photon.

[3] In this, way we emphasize the particular case of Schrödinger covariance states despite of general grout of Schrödinger minimum uncertainty states, minimizing more precise the unequality (6), so the coherent and squeezed minimize it.

When the state of quantum system is with zero covariance of A and B, than Schrödinger relation becomes the Heisenberg inequality. In this sense it is more general (and more precise), as Schrödinger mentioned in his article. This is an advantage of Schrödinger uncertainty relation compared to the Heisenberg one [5, 6].

Actually, the Schrödinger's paper is mainly based on the notes of the seminars of Physics-Mathematical section of the Prussian Academy, where many famous physicists worked to establish the underlying basis of Quantum Theory. Being a kind of internal report, this work remained at certain marginal distance for many years from the physicist scientific awareness, and no book on quantum mechanics has quoted it. Having in mind our goal – to make the article more popular, we have translated the original article of Schrödinger in English [6]. Schrödinger's paper, originally written in German, was translated in Russian by Rogali (1976) [24] (and in Polish but the information here was not confirmed).

Another argument in favor of its oblivion concerns the enthusiastic discussions, mostly about the physical interpretation of the uncertainty principle, rather than its mathematical straightforward derivation. After Schrödinger (1930) and Robertson (1930, 1934) the first appearance of this new uncertainty relation is in the Merzbacher's book [18]. However, he did not pay enough attention to the new term (covariance) and directly derived the Heisenberg relation. During 1980 in [22] the authors discussed the relation and there it was generalized for the case of non-Hermitian operators and for mixed states also. Few years later (1989), de la Torre (and collaborators) from University of de Mar del Plata [30], pay special attention to the exact uncertainty relation of Schrödinger. In their work, on the base of Quantum Covariance Function (QCF), the uncertainty and nonseparability are discussed in details and reformulation of the original EPR argument [31] is considered too. The complex second order expectation value QCF in [30] is constructed in such way that its real part is exactly the well known covariance [8] and the imaginary part is quantum mean value of the commutator. To avoid misunderstanding it is worth noting the difference with our terminology, where covariance states are functions (vectors) of the Hilbert space $\mathcal{H}$, minimizing the Schrödinger uncertainty relation (6) with non-zero covariance.

Circumstantial proof[4] and detailed analysis of the covariance, done by Schrödinger, are strong reasons to call the states with non-vanishing covariance - "Schrödinger covariance states". In this way, we emphasize only one of the three possible classes, minimizing the uncertainty relation - the states with Cov (A, B) ≠ 0 (the other two are coherent and squeezed). We think that it is more correct to call them covariance, than correlated, since the new term added by Schrödinger (1930) is exactly the covariance[5].

Significant contribution to this topic gives D. Trifonov, who makes generalizations of the uncertainty relation in several directions [26-29] and he is one of the main creators of the theory of linear invariants (first integrals of equations of motion) for arbitrary time-dependent quadratic systems in Quantum Mechanics [25]. This method gives the possibility to express the time-evolutions of the two variances $(\Delta A)^2 \equiv Var(A;t)$ and $(\Delta B)^2 \equiv Var(B;t)$ for such arbitrary quantum systems in explicit forms [25, the third article]. The time-evolutions of the third independent statistical moment of second order $Cov(A,B;t)$ in general case was derived significantly later [32] and thus the method of linear invariants was completed.

---

[4] Robertson's article [7] is only 9 lines, without proof!
[5] Mathematical notion correlation, which Manko and colaborators (1980) [23] suggest, is defined on the base of notion covariance (see mathematical handbook of G. Korn and T. Korn [8]).

In 1998 D. Trifonov and S. Donev [26] (see also [28]) established new **n** relations, which they call characteristic

(10) $$C_r^{(n)}(\sigma(X;\rho)) \geq C_r^{(n)}(C(X;\rho)),$$

where $\sigma(X;\rho)$ is the covariance matrix [8], and $C(X)$ is $n \times n$ matrix of the mean value of the commutators $[X_j, X_k]$, multiplied by factor $-\frac{i}{2}$,

(11) $$\sigma(X;\rho) = \begin{pmatrix} \sigma_{11} & \cdots & \sigma_{1n} \\ \cdots & \cdots & \cdots \\ \sigma_{n1} & \cdots & \sigma_{nn} \end{pmatrix}, \quad \sigma_{jk} = \left\langle \frac{X_j + X_k}{2} \right\rangle - \langle X_j \rangle \langle X_k \rangle \text{ in state } \rho$$

(11a) $$C(X;\rho) = \begin{pmatrix} C_{11} & \cdots & C_{1n} \\ \cdots & \cdots & \cdots \\ C_{n1} & \cdots & C_{nn} \end{pmatrix}, \quad C_{jk} = -\frac{i}{2} \langle [X_j, X_k] \rangle \text{ in state } \rho.$$

$C_r^{(n)}(\Phi)$ denotes the coefficients in front of the powers of $\lambda$ in the characteristic polynomial

(12) $$\det(\Phi - \lambda I) \equiv |\Phi - \lambda I| \equiv \lambda^n + C_1^{(n)}\lambda^{n-1} + C_2^{(n)}\lambda^{n-2} + \ldots + C_{n-1}^{(n)}\lambda^1 + C_n^{(n)}\lambda^0,$$

and we will call them characteristic coefficients of the polynomial. It is the same for all similar of $\Phi$ matrices ($\Phi` = T\Phi T^{-1}$, $\Phi$ and T – ordinary). It could be shown that characteristic coefficients are sum of all main minors of $r$-th order, and $r$ is the power of $\lambda$

(13) $$C_r^{(n)}(\Phi) = \sum_{1 \leq i_1 \leq i_2 \ldots n} \begin{vmatrix} \Phi_{i_1 i_1} & \cdots & \Phi_{i_1 i_r} \\ \cdots & \cdots & \cdots \\ \Phi_{i_r i_1} & \cdots & \Phi_{i_r i_r} \end{vmatrix}.$$

The matrices, similar to each other - $\Phi$, $\Phi`$, $\Phi``$ etc., are the same linear transformation of vectors in different bases ($y = \Phi x$), but the characteristic polynomial, and consequently its coefficients $C_r^{(n)}$ stay unchanged for that transformation [11]. They differ from characteristic values ($\equiv$ eigenvalues, roots of the equation $|\Phi - \lambda I| = 0$) of the matrix $\Phi$. This property is used to generalize the uncertainty relations [26]. These invariant coefficients are number **n,** for example when $r = 1$, $C_1^{(n)}(\Phi) = Tr(\Phi)$, and when $r = n$, $C_n^{(n)}(\Phi) = \det(\Phi)$. The Schrödinger's case is $n = 2$.

Another class of uncertainty relations concerning positive definite $2N \times 2N$ covariance matrixes (consisted of trace-class invariant coefficients) for 2N observables are established in [26]

(14) $$Tr(i\sigma(X_1, X_2, \ldots, X_{2N}; \rho)J)^{2k} = 2^{1-2k} \sum_j^N \left| \langle [X'_j, X'_{N+j}] \rangle \right|^{2k},$$

where $J = \begin{pmatrix} 0 & -1 \\ 1 & 0 \end{pmatrix}$ $X'_j = \Lambda(\rho)_{jl} X_j$,

and $\Lambda$ is the simplectic matrix, which diagonalizes $\sigma(X_1, X_2, \ldots, X_{2N}; \rho)$.

We will consider in details a special case of characteristic relations (10). The case n = 3, r = 2 is presented in [26]. We will consider n = 2. When r=2

$$C_2^2(\sigma(X;\rho)) = \det\begin{vmatrix} (\Delta X_1;\rho)^2 & Cov(X_1,X_2;\rho) \\ Cov(X_2,X_1;\rho) & (\Delta X_2;\rho)^2 \end{vmatrix} = (\Delta X_1;\rho)^2(\Delta X_2;\rho)^2 - Cov^2(X_1,X_2;\rho)$$

and the inequality (10) becomes that of Schrödinger (3). When r = 1, we get:

(15) $\quad C_1^2(\sigma(X;\rho)) = \det|(\Delta X_1;\rho)^2| + \det|(\Delta X_2;\rho)^2| = (\Delta X_1;\rho)^2 + (\Delta X_2;\rho)^2$,

i.e. the sum of determinant of first order on the main diagonal and the inequalities (10) become

(16) $\quad (\Delta X_1;\rho)^2 + (\Delta X_2;\rho)^2 \geq 0$,

The Planck's constant on the right side disappear

(17) $\quad m\omega.(\Delta q)^2 + \dfrac{(\Delta p)^2}{m\omega} \geq 0$.

Now we will generalize the inequality above, keeping the commutator respectively – the canonical form. Let us consider the following inequality, which is based on the fact, that $(a - b)^2 \geq 0$ for arbitrary $a, b \in R$

(18) $\quad (\Delta A)^2 + (\Delta B)^2 \geq 2(\Delta A)(\Delta B)$,

where we replace $a = \Delta A$ and $b = \Delta B$. Appling Schrödinger uncertainty relation (3), we get for the sum of variances

(19) $\quad (\Delta A)^2 + (\Delta B)^2 \geq \sqrt{|\langle[A,B]\rangle|^2 + 4Cov^2(A,B)}$.

If we neglect $Cov^2(A, B)$ (which is always positive) we get for canonical variables $q$ and $p$, as in [29a]

(20) $\quad m\omega.(\Delta q)^2 + \dfrac{(\Delta p)^2}{m\omega} \geq \hbar$.

The relation (19) precisely specifies and generalizes (16) and (20). The two inequalities above are independent from the characteristic relations (10) and (14), and sometimes the classical case $\Delta A^2 + \Delta B^2 \geq 0$ is used in handbooks on experimental physics [10].

Now we will mention another inequality [28, 29], concerning statistical correlation between several observables in two and more states. It is invariant generalization of Schrödinger's relation, and for two states it becomes:

(21) $\quad \dfrac{1}{2}\left[(\Delta_\psi q)^2(\Delta_\varphi p)^2 + (\Delta_\varphi q)^2(\Delta_\psi p)^2\right] - |Cov(q,p;\psi)Cov(q,p;\varphi)| \geq \dfrac{\hbar^2}{4}$.

When $\psi = \varphi$ the above relation transforms into the Schrödinger uncertainty relation (3).
Since the relation (21) is neither the sum nor the product of two Schrödinger inequalities it is not trivial (it can not be presented by a sum or a product of two values dependent only on ψ and φ). This relation is a special case of type (2, 2) introduced in [29]: "State extended Uncertainty Relations of type (n, m)", n – inequalities, m – states. In that work, one can find thorough classification of many inequalities, including those ones based on the modern 'entanglement' states from the theory of quantum computers.

## Conclusion

The short review above discusses the evolution of the uncertainty relations in quantum physics and was presented at Physics Symposium dedicated to G. Nadjakoff. His investigations in the field of **internal photoeffect** (started 1925) are pioneer works in the experimental Quantum physics, leaded later to the invention of the photocopier (1937).

D. Trifonov works are one continuation of this tradition in Bulgarian Academy of Sciences. One independent opinion about method of linear invariants is that of the Nobel Prize winner Roy Glauber, who said: "I am greatly indebted to … Vladimir Man'ko for telling me many new things about harmonic oscillator" [33]. We would like to emphasize that one of the main creators of this method [25] is also D. Trifonov, which can be seen by the fact that he is a permanent collaborator of Manko's early publications.

## References

1. **W. Heisenberg**, *Über den anschaulichen Inhalt der quantentheoretishen Kinematik und Mechanik*, ZS für Physik, **43** (1927) ss. 172-198;
2. **E. Kennard**, *Zur Quantenmechanik einfacher Bewegungstypen*, ZS für Physik, **44** (1927) ss. 326-352; **H. Weyl**, *Gruppentheorie und Quantenmechanik*, Hirzel, Leipzig (1928);
3. **H. Robertson**, *The uncertainty principle*, Phys. Review, **34** (1929) pp. 163-164;
4. **R. Dichburn**, Proc. Royal Irish Academy, 39 (1932) p. 73;
5. **E. Schrödinger**, *Zum Heisenbergschen Unschärfeprinzip*, Sitzungberichten der Preussischen Akademie der Wissenschaften (Phys.-Math. Klasse), **19** (1932) ss. 296-323;
6. **A. Agelow, M. Batoni**, *Translation with annotation of the original paper of Erwin Schrödinger (1930) in English,* Bulg. J. Physics, **26**, no. 5/6 (1999) pp. 193-203, http://arxiv.org/abs/quant-ph/9903100
7. **H. Robertson**, *A general formulation of the uncertainty principle and its classical interpretation,*, Phys. Rev. **35** (1930) Abstract 40, p.667;
8. **C. Gardiner**, *Handbook on Stochastic Methods*, Springer-Verlag, Heidelberg (1983) (For definition of covariance matrix, see p. 32. In present article we take into account that the quantum variables don't commute, i.e. instead $X_i X_j$ we have used half of anticommutator $\frac{\hat{X}_i \hat{X}_j + \hat{X}_j \hat{X}_i}{2}$ );
   **G. Korn, T. Korn**, *Mathematical Handbook*, McGraw-HillBookComp.Inc., New York (1961) eq. (18.4-10);
9. **M. de Gosson**, *Symplectic Geometry and Quantum Mechanics*, Series: Operator Theory: Advances and Applications, Subseries: Advances in Partial Differential Equations, vol.166, Birkhäuser (2006) p.239;
10. **M.Andreev, V. Ludskanov**, *Laboratory Physics,* Sofia (1970) (in Bulgarian);
11. **R. Bellman**, Introduction to Matrix Analysis, McGraw-Hill Comp., Inc., New York (1960) Chapter 7.8, example 2;
12. **P. Busch, M. Grabowski, P. Lahti**, *Operational Quantum Physics*, Springer, Heidelberg (1995); **John von Neumann**, *Mathematical Foundations of Quantum Mechanics*, (1932) ;
13. **A. Angelow**, *Schrödinger Covariance States in Anisotropic Waveguides,* Report at "*Winter College on Quantum Optics*", ICTP, Trieste, Italy, 14.02-04.03.1994;
    **A. Angelow, D. Trifonov**, *Schrödinger Covariance States in Anisotropic Waveguides,* ICTP (1995) Preprint 44;
14. **K. Blow**, Invited Lecture at "*Winter College on Quantum Optics*", ICTP, Trieste, Italy, 14.02-04.03.1994;
15. **A. Angelow**, *Covariance, Squeezed and Coherent States: Proposal for Experimental Realization of Covariance State*, AIP Conference Proceedings 899 Sixth International Conference of the Balkan Physical Union, Kostantinopol, Turkey, 22-26 August 2006**,** ISBN 978-0-7354-0404-5 (2007) pp. 325-326;
16. **Carlton M. Caves**, *Quantum-mechanical noise in an interferometer*, Phys. Review, D23 (1981) p. 1693;
17. **R. Loudon and P. Knight**, *Squeezed light*, J. Mod. Optics, **34** (1987) no. 6-7, pp. 709-759 (review);
18. **Merzbacher**, *Quantum Mechanics*, 3[rd] ed. (1998) John Wiley & Sons, Inc., eq.(10.58); the situation with this relation is the same in the 1[st] (1961) and 2[nd] editions,

19. **D. Crouch**, Chapter 17 in the book, A. Yariv, Quantum Electronics, 3-rd ed. (1988) John Wiley and Sons, Inc. (in this Chapter is completely considered quantum mechanical treatment of laser beam, in contrast to semiclassical in the rest book );
20. **D. Trifonov**, J.Math. Phys. **35** (1994) p. 2297;
21. **H. Yuen**, Phys. Rev. A, **13** (1976) no. 6, pp. 2226-2243;
22. **V. Dodonov, E. Kurmishev, V. Manko**, *Generalized uncertainty relation and correlated coherent states*, Phys. Letters A, **79** (1980) no. 2-3, pp. 150–152;
23. **V. Manko** et. al., Proceedings of the Lebedev Physics Institute, Russian Academy of Sciences, Moscow, **183** (1987)**, 191** (1989);
24. **A. Rogali,** *E. Schrödinger, Collected papers on quantum mechanics*, Nauka, Moscow (1976) pp. 210-217;
25. **I. Malkin, V. Manko and D. Trifonov**, Phys. Rev. D, **2** (1970) pp. 1371-1385;
    **I. Malkin, V. Manko and D. Trifonov**, J. Math. Phys., **14** (1973) pp. 576-582;
    **D. Trifonov,** Bulg. J. Physics, **2**, no. 4 (1975) pp. 303-311;
26. **D. Trifonov and S. Donev***, Characteristic UR's*, J. Phys. A: Math. Gen. **31** (1998) 8041-8041;
    **D. Trifonov**, *Remarks on the extended characteristic UR's*, J. Phys. A: Math. Gen., (2001) **34**, L75-L78;
27. **D. Trifonov**, *Robertson inteligent states*, J. Phys. A: Math. Gen. (1997) **32**, p. 5941;
28. **D. Trifonov, S. Donev**, *State extended Uncertainty Relations*, J. Phys. A, **33** (2000) L299;
29. **D. Trifonov**, JOSA A (2000) **17**, Issue 12, pp. 2486-2495;
29a. **D. Trifonov,** The World of Physics, no.2 (2001) pp. 107-116 (in Bulgarian);
    http://arxiv.org/abs/physics/0105035 (in English);
30. **de la Torre**, **P. Catuogno, S. Ferrando**, Uncertainty and Nonseparability,
    *Found. Phys. Letters,* **2**, no. 3 (1989) pp.235-244;
31. **A. Einstein, B. Podolsky, N. Rosen**, Phys. Rev. **47** (1935) pp.777-780;
32. **A. Angelow**, Physica A, **256** (1998) pp. 485-498;
33. **R. Glauber**, *Quantum theory of particle trapping by oscillating fields*, NATO Series B, Quantum Measurements in Optics, Ed. P.Tombesi, D. F. Walls (1992) Plenum Press, New York, p.13;